\documentclass[aps,pre,twocolumn,showpacs,preprintnumbers]{revtex4}
\usepackage{times,mathptmx}
\usepackage{epsfig}
\usepackage{verbatim}
\usepackage{pstricks}

\begin{document}

\title{Extended hydrodynamics from Enskog's equation 
for a two-dimensional system \\
general formalism }

\author{Hideaki Ugawa 
and Patricio Cordero}
\affiliation{Departamento de F\'{\i}sica, FCFM, Universidad de Chile, 
Santiago, Chile \\
(Dated: \today)}

\begin{abstract}

Balance equations are derived from Enskog's kinetic equation for a
two-dimensional system of hard disks using Grad's moment expansion method.
This set of equations constitute an extended hydrodynamics for moderately
dense bi-dimensional fluids. The set of independent hydrodynamic fields in
the present formulations are: density, velocity, temperature {\em and
also}---following Grad's original idea---the symmetric and traceless
pressure tensor $p_{ij}$ and the heat flux vector $\mathbf q^{k}$. An
approximation scheme similar in spirit to one made by Grad in his original
work is made. Once the hydrodynamics is derived it is used to discuss the
nature of a simple one-dimensional heat conduction problem.  
It is shown that, not too far from equilibrium, the nonequilibrium pressure 
in this case  only depends on the density, temperature and heat flux vector.

\end{abstract}

\maketitle

\section{Introduction}

In 1921 Enskog introduced a kinetic  theory for dense gases~\cite{enskog}
which is known to yield a very good approximate description of the behavior
of gases, particularly transport phenomena, as in~\cite{resibois} and
references therein. Today Enskog's original theory is known as the standard
Enskog theory  (SET)~\cite{chapman,ferziger,resibois} because after the
pioneer work of van Beijeren and Ernst~\cite{VB} there are several new
versions of Enskog's theory collectively called revised Enskog's theory
 (RET)~\cite{bellomo}.  Among the latter there are versions that have been
extended to describe condensed matter~\cite{kirkpatrick}. To
Navier-Stokes level both SET and RET lead to the same
results~\cite{VB,garzo}, whether or not an external force is present. For
SET and RET, using approximations in section \ref{sec:simplify}, 
the same hydrodynamic equations are obtained.

Once a kinetic equation is given it is possible to derive hydrodynamic like
equations, thus reducing the degrees of freedom from those of the
velocity distribution function $f(\mathbf r, \mathbf c, t)$ to the
degrees of freedom of a finite set of hydrodynamic fields. 
Of these methods we mention the Chapman-Enskog
method~\cite{oc,oe,chapman} and Grad's moment expansion
method~\cite{grad,grad58}.  The first is widely used and 
illustrated even in textbooks. Much less attention has been paid to Grad's
method which is our central tool in this paper.  Both
Chapman-Enskog and Grad's methods are truncated expansions. 

Perhaps the first effort to obtaining hydrodynamic equations (balance
equations) from Enskog's theory using Grad's moment expansion methods was by
Schmidt, et al~\cite{hess} in 1981.  They did not start from the local
distribution function but from a Maxwell distribution function
as kernel, totally linearized their hydrodynamic equations, that is,
completely eliminated the products of differences between the hydrodynamic
variables associated to the equilibrium state and the inhomogeneous
nonequilibrium state. From this they obtained approximate hydrodynamic
equations such that the conservation laws for mass, moment and energy are
not exactly satisfied. They also obtained expressions for the transport
coefficients (viscosity and heat conductivity) and the thermal pressure.

In 1988, Kremer and Rosa~\cite{kremer} obtained  hydrodynamic equations from
the local equilibrium distribution function as kernel linearizing the
collision integral in Enskog's equation. In particular they introduced an
approximation which completely eliminates the second order terms of the
collision integral. They derived sound dispersion relations for monoatomic
gases using  normal mode analysis.

In 1991, Marques and Kremer~\cite{marques} developed Kremer and Rosa'
work~\cite{kremer} and obtained linearized hydrodynamic equations involving
the second order terms of the collision integral.  Consequently their
equations are less approximate than in Ref.~\cite{hess} although in the
bi-dimensional case, eliminating the products between the hydrodynamic
variables which are deviations from the value at equilibrium, their
equations coincide with our results.  Furthermore they obtained linearized
Burnett equations for monoatomic gases where they eliminate third order
terms from the collision integral.

In 1996, Rangel-Huerta and Velasco~\cite{velasco96} obtained the
hydrodynamic equations from the local equilibrium distribution function as
kernel with partial linearization of the collision integral in Enskog's
equation. They also eliminated collision terms as in Ref.~\cite{kremer} and
obtained the extended equations as in Ref.~\cite{marques}: in the second
cited paper of Ref.~\cite{velasco96} Grad's distribution function without
approximation is used, whereas in Ref.~\cite{marques} the authors use a
Maxwellian distribution function to get the second order collision term.
Therefore, the terms of the second order spatial derivative of the pressure
tensor and heat flow vector are added in the balance equations for the
pressure tensor and heat flow vector.  They derived generalized transport
coefficients with the methods used widely in generalized
hydrodynamics~\cite{boon}. In 1997, Ref.~\cite{lutskol} develops the
extension of Grad's moment method for the RET, and in 1998, obtained
hydrodynamic equations which are applicable only to a system undergoing
uniform shear flow~\cite{lutsko}. In 2001, Rangel-Huerta and
Velasco~\cite{velasco01} eliminated the second order terms of the spatial
derivative of the pressure tensor and heat flow vector from the balance
equations for the pressure tensor and heat flow vector of the former
results~\cite{velasco96} and obtained a similar but extended hydrodynamic
equations similar to Ref.~\cite{marques} nevertheless,
get  better agreement with molecular dynamic simulations~\cite{das}.

The present article provides extended hydrodynamic
equations derived from Enskog's equation using Grad's
moment expansion method in the bi-dimensional case. They
are more complete than a linear approximation but still
they are the result of an approximation scheme that we
later explain. As far as we know this is the first time
that Grad's method used to obtain extended
hydrodynmaic equations from Enskog's equation has been
published beyond the a linear approximation in two or three
dimensions.

In addition, we will apply this hydrodynamic equations to a one
dimensional steady state heat conduction case.  There are many studies on
this theme: for example, the experimental investigation in
Ref.~\cite{teagan} and the theoretical one in Refs.~\cite{ohwada1,ohwada2}. 
Gross and Ziering~\cite{gross} investigated it starting from Boltzmann's
equation with Grad's moment method.  To our knowledge, no one has
investigated it starting from Enskog's equation using Grad's moment method. 
Recently Kim and Hayakawa~\cite{kim} investigated it starting from
Boltzmann's equation using the Chapman-Enskog method for hard core 
particles, Maxwellian ones and BGK's ones~\cite{BGK} 
and Hayakawa et al~\cite{md} investigated it for hard
disks using molecular-dynamic simulations.  Furthermore, they tried a test
of the nonequilibrium steady state thermodynamics (SST) proposed by Sasa and
Tasaki~\cite{sasa} and criticized it. The state defined in Ref.\cite{sasa}
for a gas in a nonequilibrium steady state is a 
one-dimensional heat
conductive configuration in contact with an equilibrium state through a
special porous wall (called the perfect $\mu$ wall~\cite{sasa}). There is a
nontrivial pressure difference between the equilibrium and nonequilibrium
state parallel to the direction of the heat flow. We have considered this
phenomenon with our hydrodynamic equations in Ref.~\cite{physa8835}.

Parenthetically we mention that it is meaningful to check, using kinetic
theory, whether one of the main assumptions
of SST~\cite{sasa}---the nonequilibrium steady state pressure---is a
functional of the density, temperature and heat flow.  We find that,
in fact, it is correct provided the system is not too far from
equilibrium. This is consistent with the range predicted in
Ref.~\cite{sasa} for this nontrivial pressure difference to appear. Note
that one of the useful merits of Grad's moment method is that one can
easily find relations between nonequilibrium thermodynamic variables.

The organization of this paper is as follows.  In Sec.~\ref{sec:expansion},
Grad's moment expansion method is introduced. Section~\ref{sec:enskog} is
about Enskog's equation. In Sec.~\ref{sec:balance}  we
formulate the macroscopic balance
equations. In Sec.~\ref{sec:simplify} the simplified collision
contribution used in this paper is constructed showing that, for the
approximations introduced here, there is no difference between the
macroscopic balance equations obtained by RET and SET.  In Sec.\ref{sec:egm}
the hydrodynamic equations that our procedure yields are presented and 
 the collision frequency coefficient 
$\chi$ used in this paper are introduced.  In Sec.~\ref{sec:heat} we apply
our hydrodynamic equations to a simple heat conduction case in which there
is a steady heat current between two parallel plates at slightly different
temperatures.  The closed expression of the pressure only
depends on the hydrodynamic fields: density, temperature and heat
vector. A final remarks on
this example are given. In Sec.~\ref{sec:summ} we summarize our results.

\section{The velocity distribution function and its expansion}\label{sec:expansion}

To define dimensionless variables $A$ from the physical
variables $\bar A$ we choose the mass of the particles to be 
the unit of mass, and the temperature is measured in energy units so that 
Boltzmann's constant is 1. Further, we
take a reference distance $L$ and a
reference temperature $T_0$ so that the dimensionless
coordinates $x$, velocities $c$ and time are such that
$$
  x = \bar x /L, \qquad  c = \bar c/\sqrt{T_0} , \qquad 
t =\frac{\sqrt{T_0}}{L}\bar t
$$
Assuming that the system is in a rectangular box of size $L_x\times
L_y$ the hydrodynamic fields can be rescaled defining
dimensionless fields as follows,
\begin{eqnarray}\label{AD}
\begin{array}{lll}
n &= \displaystyle \frac{L_xL_y}{N}{\bar n} &\textrm{number density},
\\
\mathbf{v}
&= \displaystyle \frac{1}{\sqrt{T_0}}\mathbf{\bar v} &\textrm{hydrodynamic velocity},
 \\
T &= \displaystyle \frac{1}{T_0}{\bar T} &\textrm{temperature},
  \\
P_{ij} &= \displaystyle\frac{L_xL_y}{NT_0}{\bar P_{ij}} \qquad&\textrm{pressure tensor},
 \\
\mathbf{q} &=\displaystyle \frac{L_xL_y}{N\,T_0^{3/2}}\,\,\mathbf{\bar q} 
                &\textrm{energy flux},
 \\
\mathbf{F} &= \displaystyle\frac{L}{T_0}\mathbf{\bar F} &\textrm{external force},
\end{array}  
\end{eqnarray}
The distribution function $\bar f$ is replaced
by a dimensionless distribution function $f$, 
$$
  f = \frac{L_xL_yT_0}{N}{\bar f}.
$$

In the following sections some auxiliary dimensionless quantities will be used
 \begin{eqnarray}\label{aKn}
\begin{array}{lll}\alpha_{y} &= \displaystyle \frac{L_x}{L}  \textrm{ }, 
 & \\
\alpha_{y} &= \displaystyle \frac{L_y}{L}  \textrm{ },
  &\\
\rho_{0} &= \displaystyle \frac{N}{L_xL_y}\frac{\pi \sigma^2}{4} 
= \frac{\pi N\delta^2}{4\alpha_{x}\alpha_{y}} \textrm{ },
  &\\
{\rm K\!n} &= \displaystyle \frac{8\sqrt{2}}{\pi}\frac{\ell}{L}\,,
  &\textrm{Knudsen number},
 \\
\delta &= \displaystyle \frac{\sigma}{L} = {\rm K\!n}\rho_{0} 
\end{array}  
\end{eqnarray}
where $\sigma$ is the disk's diameter, $N$ is the number of
disks and { $\ell=\frac{\pi \sigma}{8\sqrt{2}\rho_{0}}$ is} 
the mean free path at equilibrium. In these units, for example, 
the free flight time for Boltzmann's gases  at equilibrium temperature $T_{0}$ 
is  $\frac{\sqrt{\pi \rm K\!n}}{8}$.
 $L$ is any macroscopic distance
that we take as relevant depending on the particular problem to
be analyzed. For the purposes of the present paper we will
typically assume that $L$ is of the order of $L_y$.

Following Grad, we make a formal eight moments expansion of the distribution
function~\cite{grad} 
\begin{equation}\label{MG}
f(\mathbf{r,c},t) = f_{M}\Phi_{G},
\end{equation}
where 
\begin{equation}\label{maxwell}
f_{M} = n(\mathbf{r},t)\left[\frac{1}{2\pi T(\mathbf{r},t)}\right]
         \exp{\left[-\frac{C\mathbf{(r},t)^{2}}{2T\mathbf{(r},t)}\right]}.
\end{equation}

In Grad's method the factor $\Phi_G$ is written as an expansion 
in Hermite polynomials of the peculiar velocity $\mathbf
C\equiv \mathbf c - \mathbf v(\mathbf r,t)$,
\begin{equation}\label{setH} 
\Phi_{G}=\sum_{n=0}^{\infty} a_{i}^{(n)}(\mathbf{r}, t)H_{i}^{(n)}(\mathbf C), 
\end{equation}
where $\mathbf c$ is the molecular velocity and where
$H_{i}^{(n)}$ is a tensor with $n$ subscripts, $i=(i_{1}
\cdots i_{n})$ as well a polynomial of 
$n$-th degree in the components of $\mathbf C$. 
Also the coefficients $a_{i}^{(n)}$ are 
tensors of order $n$. The eight moments expansion
corresponds to truncating the expansion in Eq.(\ref{setH})
up to $n=3$. In dimension 3 this leads be the well known 
13 moments methods used by Grad himself working upong
Boltzmann's equation. It is widely used.  

Grad's method consists of first replacing $f$ given by
Eq.(\ref{MG}) but with the truncated expresion for
$\Phi_G$,  in the kinetic equation (Boltzmann's or
Enskog's), and then projecting the kinetic equation to the
different Hermite polynomials used in the tructated
expansion. This procedure---deviced by Grad---leads to
obtaining as many balance equations as polynomials were
included in the truncated expansion for
$\Phi_G$.

The Hermite polynomials satisfy the orthogonality relations
\begin{equation}\label{orthogonal}
\frac{1}{2\pi T}\int_{-\infty}^{\infty}H_{a}^{m}H_{b}^{n}\exp{\left[-\frac{C^2}{2T}\right]}
 d\mathbf{C}=n!\delta_{mn}\delta_{ab},
\end{equation}
and they are, for example,  
\begin{eqnarray}
H^{0}&=&1,
\\
H_{i}^{1}&=&\frac{C_{i}}{\sqrt{T}},
\\
H_{ij}^{2}&=&\frac{C_{i}C_{j}}{T}-\delta_{ij},
\\
H_{ijk}^{3}&=& \frac{C_{i}C_{j}C_{k}}{T^{\frac{3}{2}}}
            -\frac{( C_{i}\delta_{jk}+C_{j}\delta_{ki}
                                     +C_{k}\delta_{ij})}{\sqrt{T}}
\end{eqnarray}
where $i,j=1,2$.   Instead of using 
the full set of third order Hermite polynomial $H_{ijk}^{3}$ 
we use---as Grad did----the corresponding contracted 
Hermite polynomials $H_{i}^{3}$ defined by
\begin{equation}\label{3HP}
H_{i}^{3}=\frac{C_{i}}{\sqrt{T}}\left(\frac{C^{2}}{T}-4\right).
\end{equation}

The coefficients $a_{i}^n$  in Eq.(\ref{setH}) can be 
expressed in terms of the hydrodynamic fields 
$n(\mathbf{r},t)$, $T(\mathbf{r},t)$, 
and the flux vectors associated with the velocity and the energy, 
called the kinetic part of pressure tensor and the energy flux:
$P^{\rm k}_{ij}(\mathbf{r},t)$,  $q^{\rm k}_i(\mathbf{r},t)$  according to the 
following sum rules,
\begin{eqnarray}\label{defc}
\int fd\mathbf{c} &=& n\mathbf{(r},t)\,, \\
\int \mathbf{C(r,t)}fd\mathbf{c} &=& 0\,, \\
\int \frac{1}{2}\mathbf{C(r,t)}^{2}fd\mathbf{c} 
         &=& n\mathbf{(r,t)}T\mathbf{(r},t)\,,\\
\int C_{i}\mathbf{(r},t)C_{j}\mathbf{(r},t)fd\mathbf{c} 
 &=& P^{\rm k}_{ij}\,, \\
\int \frac{1}{2}\mathbf{C(r,t)}C^{2}\mathbf{(r},t)f
d\mathbf{c} &=& \mathbf{q}^{\rm k}(\mathbf{r},t),
\end{eqnarray}
where  $p_{ij}$ is the traceless part of $P_{ij}^k$,
$$
P^{\rm k}_{ij} = n(\mathbf{r},t)T(\mathbf{r},t)\delta_{ij} 
                                 + p_{ij}(\mathbf{r},t)\,.
$$  

\smallskip

The factor $\Phi_{G}$ for the eight moments method turns out to be
\begin{equation}\label{PhiG}
\Phi_{G} = 1 + \frac{1}{2n\,T^2}p_{i,j}\, C_iC_j 
             + \frac{1}{ n\,T^2}
               \left[\frac{C^2}{4T}-1\right]\mathbf{C\cdot q}^{k}
\end{equation}

We refer to the ``kinetic part''  of the fluxes to stress
that kinetic theory, in principle, includes
contributions to the fluxes associated to the interaction between
the particles.

\section{Enskog's equation}\label{sec:enskog}

Enskog's equation is 
\begin{widetext}
\begin{eqnarray}  \label{eqEnsk}
{[\frac{\partial}{\partial t} 
      + \mathbf{c_{1}}\cdot\nabla_{1} 
      + \mathbf{F\cdot \nabla_{c_{1}}}      
]f(\mathbf{r_{1},c_{1}},t)}
  &=& \frac{4}{\pi \rm K\!n}\int 
   [\chi(\mathbf{r_{1}}, \mathbf{r_{1}} + \delta \mathbf{k} \,|\, n)
      f(\mathbf{r_{1},c_{1}'},t)
      f(\mathbf{r_{1}} + \delta \mathbf{k},\mathbf{c_{2}'},t)
\nonumber \\
      &&-\chi(\mathbf{r_{1}}, \mathbf{r_{1}} - \delta \mathbf{k}\,|\, n)
      f(\mathbf{r_{1}},\mathbf{c_{1}},t)
      f(\mathbf{r_{1}} - \delta \mathbf{k},\mathbf{c_{2}},t)]
      (\mathbf{g\cdot k)}\,\theta_{\mathbf{k}}\,d\mathbf{k}\,d\mathbf{c_2}
\end{eqnarray}
\end{widetext}
where $\mathbf{k}$ is the unit vector  from the disk centered at
$\mathbf{r}_1+\delta\mathbf{k}$ to the disk with center at
$\mathbf{r}_1$ upon collision and is is integrated over all
the unit vectors while $\mathbf{g} = \mathbf{c}_2-\mathbf{c}_1$ 
and $\theta_{\mathbf{k}}$ is the Heaviside function
$\theta_{\mathbf{k}}=\theta(\mathbf{g\cdot k})$ and where
the two velocities of the disks after collision are
$\mathbf{c}_{1}' = \mathbf{c}_1 +
(\mathbf{g}\cdot\mathbf{k}) \mathbf{k}$ and
$\mathbf{c}_{2}' = \mathbf{c}_2 -
(\mathbf{g}\cdot\mathbf{k}) \mathbf{k}$, respectively.

The collision frequency 
$\chi(\mathbf{r_{1}}, \mathbf{r_{1}}\pm \delta \mathbf{k}\,|\,n)$ is the pair 
distribution function of two hard disks at contact 
and $n$ is the number density defined by Eq. (\ref{defc}) . 
This $\chi$ is given by the procedure of statistical mechanics, that is, 
by the virial expansions of the pair distribution function 
and of course its expression depends on whether one uses SET or RET.

Concretely, with $\mathbf{r}'=\mathbf{r}+\delta\mathbf{k}$, 
\begin{widetext}
\begin{eqnarray}\label{chies}
 \chi^{SET}(\mathbf{r},\mathbf{r}'|n) &=&  
1 + n(\frac{1}{2}(\mathbf{r}+\mathbf{r}'))\int V(\mathbf{r}|\mathbf{r}_{3})d\mathbf{r}_3
+ \frac{1}{2!}n^{2}(\frac{1}{2}(\mathbf{r}
  +\mathbf{r}'))\int V(\mathbf{r},\mathbf{r}'|\mathbf{r}_3\mathbf{r}_4)
d\mathbf{r}_3d\mathbf{r}_4+\cdots 
\\ \nonumber
 &=&\chi_{c}+\frac{\delta}{2}\mathbf{k}\cdot \nabla \chi_{c} 
+ \frac{\delta^{2}}{8}\mathbf{k}\mathbf{k}:\nabla \nabla \chi_{c} +\cdots
\\
\chi^{RET}(\mathbf{r},\mathbf{r}'|n) &=&  1 
+ \int n(\mathbf{r}_3)V(\mathbf{r},\mathbf{r}'|\mathbf{r}_{3})d\mathbf{r}_3
+ \frac{1}{2!}\int n(\mathbf{r}_3)n(\mathbf{r}_4)V(\mathbf{r},\mathbf{r}'|\mathbf{r}_3\mathbf{r}_4)
d\mathbf{r}_3d\mathbf{r}_4+\cdots 
\\ \nonumber
&=&\chi_{c}+\frac{\delta}{2}\frac{\partial \chi_{c}}{\partial n}(\mathbf{k}\cdot \nabla n) 
+ \frac{\delta^{2}}{8}\frac{\partial^{2}\chi_{c}}{\partial n^{2}}(\mathbf{k}\cdot\nabla n)
(\mathbf{k}\cdot\nabla n) +\cdots
\end{eqnarray}
where
\begin{equation}\chi_{c}=1+n(\mathbf{r})\int V(\mathbf{r}, \mathbf{r}'|\mathbf{r}_{3})d\mathbf{r}_{3}
              +\frac{n^{2}(\mathbf{r})}{2!}\int V(\mathbf{r}, \mathbf{r}'|\mathbf{r}_{3}\mathbf{r}_{4})
              d\mathbf{r}_{3}d\mathbf{r}_{4} + \cdots
\end{equation}
\end{widetext}
and where the usual Hushimi V-function is introduced~\cite{garzo,
uhlenbeck}.

Expanding the right hand side in powers of $\delta$ up to second order 
following the Appendix of { ~\cite{garzo},}
{
\begin{eqnarray}\label{expanJ}
 {\rm rhs^{SET}} &=&  J_{0}^{SET} + \delta J_{1}^{SET}
                                  + \delta^{2}J_{2}^{SET}, 
\\
 {\rm rhs^{SET}} &=&  J_{0}^{RET} + \delta J_{1}^{RET} 
                                  + \delta^{2}J_{2}^{RET},            
\end{eqnarray}}
where {
\begin{eqnarray}\label{SETRET}
J_{0}^{SET}&=&J_{0}^{RET}=J_{0},
\nonumber \\
J_{1}^{SET}&=&J_{1}^{RET}=J_{11}+J_{12},
\nonumber \\
J_{2}^{SET}&=&J_{21}+J_{22}+J_{23},
\nonumber \\
J_{2}^{RET}&=&J_{2}^{SET}+J_{24},
\end{eqnarray}}
{\small and
\begin{equation} \label{taylor}
\begin{array}{ll}
J_{0} &= \frac{4}{\pi \rm K\!n}\chi_{c}\int (f_{1}'f_{2}'- f_{1}f_{2})
        (\mathbf{g\cdot k})\theta_{\mathbf{k}}d\mathbf{k}d\mathbf{c_{2}},
\\ & \\
J_{11} &= \frac{4}{\pi \rm K\!n}\chi_{c}\int \mathbf{k}\cdot 
 (f_{1}'\nabla f_{2}'+ f_{1}\nabla f_{2})
        (\mathbf{g\cdot k})\theta_{\mathbf{k}}d\mathbf{k}d\mathbf{c_{2}},
 \\ & \\
J_{12} &= \frac{2}{\pi \rm K\!n}\int  
(\mathbf{k}\cdot \nabla \chi_{c})(f_{1}'f_{2}'+f_{1}f_{2})
        (\mathbf{g\cdot k})\theta_{\mathbf{k}}d\mathbf{k}d\mathbf{c_{2}},
\\ & \\
J_{21} &= \frac{2}{\pi \rm K\!n}\chi_{c}\int  
\mathbf{kk}:(f_{1}'\nabla \nabla f_{2}'-f_{1}\nabla \nabla f_{2})
        (\mathbf{g\cdot k})\theta_{\mathbf{k}}d\mathbf{k}d\mathbf{c_{2}},
\\ & \\
J_{22} &= \frac{2}{\pi \rm K\!n}\int  
(\mathbf{k}\cdot \nabla \chi_{c})\mathbf{k}
(f_{1}'\nabla f_{2}'-f_{1}\nabla f_{2})
        (\mathbf{g\cdot k})\theta_{\mathbf{k}}d\mathbf{k}d\mathbf{c_{2}},
\\ & \\
J_{23} &= \frac{1}{2\pi \rm K\!n}\int  
(\mathbf{kk}:\nabla \nabla \chi_{c})(f_{1}'f_{2}'-f_{1}f_{2})
        (\mathbf{g\cdot k})\theta_{\mathbf{k}}d\mathbf{k}d\mathbf{c_{2}},
\\ & \\
J_{24} &= -\frac{1}{2\pi \rm K\!n}\frac{\partial \chi_{c}}{\partial n}
\int (\mathbf{kk}:\nabla \nabla n(\mathbf{r},t))(f_{1}'f_{2}'- f_{1}f_{2})
        (\mathbf{g\cdot k})\theta_{\mathbf{k}}d\mathbf{k}d\mathbf{c_{2}}
\end{array}
\end{equation}
}
\noindent where the following abbreviations: $f_{1}=f(\mathbf{r_{1}, 
c_{1}}, t)$, $f_{1}'=f(\mathbf{r_{1}, c_{1}'}, t)$, $f_{2}=f(\mathbf{r_{1}, 
c_{2}}, t)$ and $f_{2}'=f(\mathbf{r_{1}, c_{2}'}, t)$ are introduced.
Here $\chi_{c}$ is  concretely given by the virial expansions~\cite{garzo},
but, as seen below, we do not use such an expression.

\section{The balance equations}\label{sec:balance}

The macroscopic balance equations are derived following Grad's
prescription, successively projecting the kinetic equation to
the different Hermite polynomials being used.

If Enskog's equation is multiplied by a function $\phi_1$ of
the peculiar velocity $\mathbf C_1 = \mathbf c_1-\mathbf v$ and
integrating over $\mathbf c_1$ yields 
\begin{eqnarray}\label{beg}
 &&\int \phi_{1} \frac{\partial}{\partial t}f_{1}d\mathbf{c_{1}} 
      + \int \phi_{1}\mathbf{c_1}\cdot \nabla f_{1}d\mathbf{c_{1}}
      + \int \phi_{1}\mathbf{F}\cdot \nabla_{c_{1}}f_{1}d\mathbf{c_{1}}
    \nonumber \\
 && =I
\nonumber \\
 && =I_{0}^{SET} + \delta I_{1}^{SET} + \delta^{2} I_{2}^{SET} 
       \qquad (\verb|for SET|)
\nonumber \\
 && =I_{0}^{RET} + \delta I_{1}^{RET} + \delta^{2} I_{2}^{RET} 
       \qquad (\verb|for RET|)
\end{eqnarray}
where
\begin{equation}\label{Ii}
    I_{i}=\int \phi_{1} J_{i}d\mathbf{c_{1}} \qquad (i=0,1,2).
\end{equation}
The balance equations of our interest, in particular the basic
conservation equations (mass, momentum and energy), are obtained
multiplying Enskog's equation by
$\phi=1, \mathbf{c_{1}}, \frac{1}{2}C_{1}^{2}$. In such cases 
$\phi_{1} + \phi_{2} = \phi_{1}' + \phi_{2}'$ is true (prime
indicates post collision velocities) so that
\begin{equation}
I_{0}^{SET} = I_{0}^{RET} = 0,
\end{equation}
and
\begin{widetext}
\begin{equation}\label{befcc1}
I_{1}^{SET} = I_{1}^{RET}
=\frac{2}{\pi \rm K\!n}\nabla \cdot \int \chi_{c}(\phi_{1}-\phi_{1}')f_{1}f_{2}
        (\mathbf{g\cdot k})\theta_{\mathbf{k}}\mathbf{k}
        d\mathbf{k}d\mathbf{c_{1}}d\mathbf{c_{2}},
\end{equation}
\begin{equation}\label{befcc2}
I_{2}^{SET} = I_{2}^{RET}= \frac{1}{\pi \rm K\!n} \nabla \cdot \int \chi_{c}(\phi_{1}-\phi_{1}')
          \left[
           \mathbf{k}\cdot f_{1}f_{2}\nabla \log{\frac{f_{1}}{f_{2}}}
                       \right]
            (\mathbf{g\cdot k})\theta_{\mathbf{k}}
            \mathbf{k}d\mathbf{k}d\mathbf{c_{1}}d\mathbf{c_{2}}.
\end{equation}
{
In the case when the external force is uniform 
the general conservation equation may be written as
\begin{equation}\label{begc}
 \frac{\partial}{\partial t}\int \phi_{1}f_{1}d\mathbf{c_{1}} 
      + \nabla \cdot\int \phi_{1}\mathbf{c_1}f_{1}d\mathbf{c_{1}}
      - \mathbf{F}\cdot \int \phi_{1}\nabla_{c_{1}}f_{1}d\mathbf{c_{1}}
  = -\nabla \cdot \Psi
\end{equation}
where $I=-\nabla \cdot \Psi$.
}

In the case when $\phi_1$ is not associated to a microscopic 
conservation law   the results are
{
\begin{equation}
I_{0}^{SET} = I_{0}^{RET} = \frac{4\chi_{c}}{\pi \rm K\!n}\int (\phi_{1}'-\phi_{1})f_{1}f_{2}
         (\mathbf{g\cdot k})\theta_{\mathbf{k}}
d\mathbf{k}d\mathbf{c_{1}}d\mathbf{c_{2}},
\end{equation}
\begin{equation}\label{ctpq1}      
I_{1}^{SET} = I_{1}^{RET}
             = \frac{4}{\pi \rm K\!n}\int (\phi_{1}-\phi_{1}')
\left[\chi_{c}\mathbf{k}\cdot f_{1}\nabla f_{2}
        +\frac{1}{2}(\mathbf{k}\cdot\nabla_{\mathbf{r}}\chi_{c})f_{1}f_{2}\right]
        (\mathbf{g\cdot k})\theta_{\mathbf{k}}d\mathbf{k}d\mathbf{c_{1}}
        d\mathbf{c_{2}},
\end{equation}
\begin{equation}\label{ctpq2s}
I_{2}^{SET} = \frac{2}{\pi \rm K\!n}\int (\phi_{1}'-\phi_{1})
\left[(\mathbf{k}\cdot \nabla\chi_{c})\mathbf{k}(f_{1}\nabla f_{2})
        +\chi_{c}\mathbf{kk}:f_{1}\nabla \nabla f_{2}
        +\frac{1}{4}(\mathbf{kk}:\nabla \nabla \chi_{c})f_{1}f_{2}
\right]
            (\mathbf{g\cdot k})\theta_{\mathbf{k}}
d\mathbf{k}d\mathbf{c_{1}}d\mathbf{c_{2}},
\end{equation} 
\begin{equation}\label{ctpq2r}
I_{2}^{RET} = \frac{2}{\pi \rm K\!n}\int (\phi_{1}'-\phi_{1})
\left[(\mathbf{k}\cdot \nabla\chi_{c})\mathbf{k}(f_{1}\nabla f_{2})
        +\chi_{c}\mathbf{kk}:f_{1}\nabla \nabla f_{2}
        +\frac{1}{4}\frac{\partial^{2}\chi_{c}}{\partial n^{2}}
(\mathbf{k}\nabla n(\mathbf{r},t))(\mathbf{k}\nabla n(\mathbf{r},t))f_{1}f_{2}
\right]
            (\mathbf{g\cdot k})\theta_{\mathbf{k}}
d\mathbf{k}d\mathbf{c_{1}}d\mathbf{c_{2}}.
\end{equation}}
\end{widetext}

\section{Simplified collision contribution}\label{sec:simplify}

The collisional term in both Boltzmann's and  Enskog's
equations is quadratic in the
distribution function, hence there are integrals which involve
products $H^\ell$\,$H^m$ which have to be projected to different
$H^n$s. Even in the case of the 13 moments method for Boltzmann's equation
with the contraction as in Eq.~(\ref{3HP}),
Grad proposed to consider only those contributions such
that $\ell+m=n$ even though there is no clear  mathematical
criterion to make such simplification  except for Maxwellian molecules  
because this case the collision term is integrable and such quadratic 
terms do not appear.

Nevertheless, Grad was able to prove that very little error
is made even in the case of non-Maxwellian molecules by
restricting the contributions to $\ell+m=n$~~\cite{grad,grad58}. 
Although we do not have such a prove in the case of
Ebskog's equation we follow the same prescription.

In our present case, to be able to obtain general
hydrodynamic equations ---following the present
procedure--- which are not overwhelmingly complex it is
necessary to make some further simplifications.

Since $\Phi_G$ is a correcting factor we may consider that its
contribution can be regarded as $1$ plus something small,
schematically  $\Phi_G = 1 + \varepsilon$. Since the collision
integral is quadratic in the distribution function, the
integrand has a product $\Phi_{G1}\Phi_{G2} = 1 + \varepsilon_1
+ \varepsilon_2 + \varepsilon_1\varepsilon_2$. Our
simplificatory strategy will be first that: to evaluate $J_0$ we keep all
the contributions; to evaluate $J_1$ we keep the ``order
$\varepsilon$'' terms, namely, $\Phi_{G1}\Phi_{G2}
\approx \Phi_{G1} + \Phi_{G2}-1$; to evaluate $J_2$ 
we take $\Phi_{G1}\Phi_{G2}\approx 1$.

Therefore, in Eq.~(\ref{ctpq1}), the following relations are required,    
$$
\begin{array}{ll}
f_{1}f_{2} &=  f_{M1}f_{M2}(\Phi_{G1}+\Phi_{G2}-1),
\\ & \\
f_{1}\nabla f_{2} &=  f_{M1}(\nabla f_{M2})(\Phi_{G1}+\Phi_{G2}-1)
                     + f_{M1}f_{M2}\nabla \Phi_{G2}                 
\end{array}
$$
and in Eqs.~(\ref{ctpq2s}) and (\ref{ctpq2r}) we use
\begin{equation}\label{app2}
f_{1}f_{2} =  f_{M1}f_{M2}.
\end{equation}
For hard disks, $f_{M1}f_{M2} =  f_{M1}'f_{M2}'$. 
Hence, $J_{24}$ in Eq.~(\ref{taylor}) vanishes, and
from Eq.~(\ref{SETRET}), the SET's collision term coincides with the RET's.   
Additionally, also $J_{23}$ in Eq.~(\ref{taylor}) vanishes 
for the same reason. 
Consequently, Eq.~(\ref{ctpq2r}) is reduced (\ref{ctpq2s}) and
\begin{widetext}
\begin{equation}\label{ctpq2}
I_{2}^{SET} = I_{2}^{RET}
            =\frac{2}{\pi \rm K\!n}\int (\phi_{1}'-\phi_{1})
\left[(\mathbf{k}\cdot \nabla\chi_{c})\mathbf{k}(f_{M1}\nabla f_{M2})
        +\chi_{c}\mathbf{kk}:f_{M1}\nabla \nabla f_{M2}\right]
            (\mathbf{g\cdot k})\theta_{\mathbf{k}}
d\mathbf{k}d\mathbf{c_{1}}d\mathbf{c_{2}}.
\end{equation}

\end{widetext}
After introducing the approximations mentioned above, we directly calculate, using
{\sc Maple} routines, the collisional term in Enskog's equation for all
contributions originated in the products $H^{l}H^{m}$.

\section{The hydrodynamic equations}\label{sec:egm}

The central result of the present paper are the hydrodynamic
equations that  we have derived from Enskog's equation according
to what has been explained in the previous sections. 
 
To the best of our knowledge, these hydrodynamic equations are published for
the first time. For the two dimensional case we have not encountered any paper
at all while for three dimensional case the published literature refers to 
linearized versions. For the sake of clarity the balance equations for conserved
quantities (mass, momentum and energy) are written in terms of the
hydrodynamic quantities: density, hydrodynamic velocity, temperature,
pressure tensor and heat flux; whereas the balance equations associated to
the pressure tensor and the heat flux are written in terms of the symmetric
pressure tensor and heat flux vector. This is what is also done in almost
all well-known articles for the extremely dilute system.

We use the following notation as Euler's (material) derivative:
$\frac{D}{Dt} = \frac{\partial}{\partial t} 
              + \mathbf{v}\cdot \frac{\partial}{\partial \mathbf{r}}$.
Summation is implied with repeated tensorial indeces.

The final extended hydrodynamic equations are the following:
\begin{equation}
 \frac{Dn}{Dt} = -n \nabla \cdot \mathbf{v}
\end{equation}
\begin{equation}\label{NSeq}
 n\frac{D\mathbf{v}}{Dt} = { n\mathbf{F}}-\nabla \cdot \mathbf{P} 
\end{equation}
\begin{equation}
 n\frac{DT}{Dt} = -(\nabla\cdot \mathbf{Q} 
                        + \mathbf{P}:\nabla \mathbf{v}) 
\end{equation}
\begin{eqnarray}
 &&\frac{D}{Dt}p_{ij} + \frac{\partial v_{l}}{\partial x_{l}}p_{ij} 
                    + p_{il}\frac{\partial v_{l}}{\partial x_{i}} 
                    + p_{jl}\frac{\partial v_{i}}{\partial x_{l}}
                    - p_{lm}\frac{\partial v_{m}}{\partial x_{l}}\delta_{ij} 
\nonumber \\
             &&\verb|    | + \frac{1}{2}
             \left[ \frac{\partial}{\partial x_{i}}q_{j}^{k}
                   +\frac{\partial}{\partial x_{j}}q_{i}^{k}
                   -\frac{\partial}{\partial x_{l}}q_{l}^{k}\delta_{ij}\right] 
                    + nTS_{ij}      
\nonumber \\
              && \verb|                            |   = \mathbf{I_{p}}
\end{eqnarray}
\begin{eqnarray}
 &&  \frac{D}{Dt}q_{i}^{k} 
   + \frac{3}{2}\left[ \frac{\partial v_{i}}{\partial x_{l}}q_{l}^{k}
                 + \frac{\partial v_{l}}{\partial x_{l}}q_{i}^{k}
                             \right]
   + \frac{1}{2}\frac{\partial v_{l}}{\partial x_{i}}q_{l}^{k}
\nonumber \\
  &&\verb|    |+ T\frac{\partial}{\partial x_{l}}p_{li}
    + 2\frac{\partial T}{\partial x_{l}}p_{li}
    - \frac{1}{n}\frac{\partial}{\partial x_{m}}P_{ml}^{k}p_{li}
\nonumber \\
   &&\verb|    |+2nT\frac{\partial T}{\partial x_{i}} 
   = \mathbf{I_{q}}                    
\end{eqnarray}
where {the pressure tensor $\mathbf{P}$ is given as
  \begin{equation}\label{defpressure}
  \mathbf{P} = \beta_{p} \mathbf{p} + \mathbf{P_{col}}, 
\end{equation}
{\small
{with
\begin{eqnarray}\label{defpcol}
 \beta_{p} &=& 1+\frac{\delta}{\rm K\!n}\chi_{c} n, 
 \\
 \mathbf{P_{col}}
 &=& \left[ 1+\delta \frac{2}{\rm K\!n}\chi_{c} n\right]nT\mathbf{1}
    - \frac{\delta^{2}}{\sqrt{\pi}\rm K\!n}\chi_{c} n^{2}\sqrt{T}
   \left[ \mathbf{S} + 2(\nabla \cdot \mathbf{v})\mathbf{1}
 \right]
\end{eqnarray}
}

\noindent and the shear tensor $\mathbf{S}$ which is defined by
\begin{equation} \label{def:S}
S_{ij}=   \frac{\partial v_{j}}{\partial x_{i}}
          +\frac{\partial v_{i}}{\partial x_{j}}
          -\frac{\partial v_{l}}{\partial x_{l}}\delta_{ij},
\end{equation}
{ where $\mathbf{Q}= \beta_{q}\mathbf{q^{k}}+\mathbf{q_{col}}$, 
\begin{equation}
\beta_{q}= 1 + \delta\frac{3}{2\rm K\!n}\chi_{c} n, \qquad
 \mathbf{q_{col}}=  
           -\delta^{2}\frac{2}{\sqrt{\pi}\rm K\!n}\chi_{c} n^{2}\sqrt{T}\nabla T
\end{equation}}                   
and where the kinetic contribution to the heat flux vector $\mathbf{Q}$ is
given by $\mathbf{q}^{k}$ following the definition Eq.
(\ref{defc}). Additionally, the remaining terms, which give the collisional 
contribution to the extended hydrodynamic equations, are given as below. 
$\mathbf{I_{p}}$  is split in the form 
$\mathbf{I_{p}}= \mathbf{I_{p}^{0}+I_{p}^{1}+I_{p}^{2}}$ 
where
\begin{equation}\label{Btermp} 
 \mathbf{I_{p}^{0}}= - \frac{8}{\sqrt{\pi}{\rm K\!n}}\chi_{c} \sqrt{T}
 \left[ np_{ij} + \frac{1}{128T^{2}}
               \left[2q_{i}^{k}q_{j}^{k}-q_{l}^{k}q_{l}^{k}\delta_{ij}\right]
 \right], 
\end{equation}
\begin{eqnarray}
\mathbf{I_{p}^{1}} 
= &-&\frac{\delta}{\rm K\!n}\left[\verb|   |
  \frac{5}{4}\chi_{c} \left[ q_{i}^{k}\frac{\partial n}{\partial x_{j}}
                         +q_{j}^{k}\frac{\partial n}{\partial x_{i}}
                         -q_{l}^{k}\frac{\partial n}{\partial x_{l}}\delta_{ij}
\right]\right.
\nonumber \\
      &&\left. \verb|    |
+\frac{3}{4}\chi_{c} \left[\frac{\partial}{\partial x_{i}}q_{j}^{k}
                    +\frac{\partial}{\partial x_{j}}q_{i}^{k}
                    -\frac{\partial}{\partial x_{l}}q_{l}^{k}\delta_{ij}
\right]\right.
\nonumber \\
      &&\left. \verb|    |
+n\left[ q_{i}^{k}\frac{\partial \chi_{c}}{\partial x_{j}}
               +q_{j}^{k}\frac{\partial \chi_{c}}{\partial x_{i}}
               -q_{l}^{k}\frac{\partial \chi_{c}}{\partial x_{l}}\delta_{ij}
\right]\right.
\nonumber \\
      &&\left. \verb|    | 
-\chi_{c} \frac{\partial v_{l}}{\partial x_{l}}p_{ij}
       +\frac{5}{4}\chi_{c} n^{2}TS_{ij}\right]
\end{eqnarray}
and the $\delta^2$ contribution is
\begin{eqnarray}
\mathbf{I_{p}^{2}} &=& \frac{\delta^{2}}{{\rm K\!n}}\frac{1}{\sqrt{\pi}}
\left[ \verb|  |2\chi_{c} n\sqrt{T}\left[ 
  \frac{\partial n}{\partial x_{i}}\frac{\partial T}{\partial x_{j}}
 +\frac{\partial n}{\partial x_{j}}\frac{\partial T}{\partial x_{i}}
 -\frac{\partial n}{\partial x_{l}}\frac{\partial T}{\partial x_{l}}\delta_{ij}
\right]\right.
\nonumber \\  
  && \left. \verb|       |+\frac{1}{2}\chi_{c} \frac{n^{2}}{\sqrt{T}}\left[
 2\frac{\partial T}{\partial x_{i}}\frac{\partial T}{\partial x_{j}}
- \frac{\partial n}{\partial x_{l}}\frac{\partial T}{\partial x_{l}}\delta_{ij}
\right]\right.
\nonumber \\
 && \left. \verb|       |+\chi_{c} n^{2}\sqrt{T}\left[
  2\frac{ \partial^{2}T}{\partial x_{i}\partial x_{j}}
         -\frac{\partial^{2}}{\partial x_{l}\partial x_{l}} T
\right]\right.
\nonumber \\ 
    &&\left. \verb|       |+ n^{2}\sqrt{T}\left[
 \frac{\partial T}{\partial x_{i}}\frac{\partial \chi{c}}{\partial x_{j}}
+\frac{\partial T}{\partial x_{j}}\frac{\partial \chi_{c}}{\partial x_{i}}
-\frac{\partial T}{\partial x_{l}}
                 \frac{\partial \chi_{c}}{\partial x_{l}}\delta_{ij}
\right]\right.
\nonumber \\
&&\left. \verb|       |
+2\chi_{c} n^{2}\sqrt{T}\frac{\partial v_{l}}{\partial x_{l}}S_{ij}\right],
\end{eqnarray}

$\mathbf{I_{q}}$  is split 
$\mathbf{I_{q}}= \mathbf{I_{q}^{0}+I_{q}^{1}+I_{q}^{2}}$ and
 \begin{eqnarray}\label{Btermq}
 \mathbf{I_{q}^{0}}= - \frac{\chi_{c}}{\sqrt{\pi T}{\rm K\!n}}
\left[ 4nTq_{i}^{k} + p_{il}q_{l}^{k}\right], 
 \end{eqnarray}
 $\mathbf{I_{q}^{1}} = \frac{\delta}{\rm K\!n}
         [(1)_{\delta} + (2)_{\delta} + (3)_{\delta} ] $ 
\begin{eqnarray}
 (1)_{\delta} &=& -\frac{1}{2}
          \left[\verb|  |7\chi_{c} nT\frac{\partial}{\partial x_{l}}p_{li} 
            +7T\chi_{c} \frac{\partial n}{\partial x_{l}}p_{li}\right.
\nonumber \\
     &&\left. \verb|    |
             +7nT\frac{\partial \chi_{c}}{\partial x_{l}}p_{li} 
             + 4\chi_{c}\frac{\partial T}{\partial x_{l}}p_{li}\right],
 \nonumber \\
 (2)_{\delta} &=& -\left[ 8\chi_{c} nT^{2}\frac{\partial n}{\partial x_{i}} 
                         +7\chi_{c} n^{2}T\frac{\partial T}{\partial x_{i}}
                         +4n^{2}T^{2}\frac{\partial \chi_{c}}{\partial x_{i}}
\right],
 \nonumber \\
 (3)_{\delta} &=& -\frac{1}{2}\chi_{c} n\left[
                  2S_{il} - \frac{\partial v_{l}}{\partial x_{l}}\delta_{ij}
                                                   \right]q_{l}^{k}
\end{eqnarray}
and where
\begin{eqnarray}
 \mathbf{I_{q}^{2}} 
&=& \frac{\delta^{2}}{8\rm K\!n\sqrt{\pi}}n\sqrt{T}
 \left[ \verb|  |35\chi_{c} n^{2}\frac{\partial T}{\partial x_{l}}S_{li}
  +18nT\frac{\partial \chi_{c}}{\partial x_{l}}S_{li}\right.
\nonumber \\ 
&&\left.\verb|          |+36\chi_{c} T\frac{\partial n}{\partial x_{l}}S_{li}
+18\chi_{c} nT\frac{\partial^{2}}{\partial x_{l}\partial x_{l}} v_{i}\right.
\nonumber \\
&&\left. \verb|          |
+36\chi_{c} nT\frac{\partial}{\partial x_{i}}\frac{\partial v_{l}}{\partial x_{l}}
\right].
\end{eqnarray}
  
The above equations have the well known form which comes from very general
arguments from continuous fluid matter and where established in the XIX
century. The detail structure of some of the terms comes from kinetic
theory.

Of course it is interesting to compare our results to the ones 
obtained by means of the Chapman-Enskog method.  
For example, Kim and Hayakawa~\cite{kim} have derived the explicit velocity  
distribution function of Bolzmann's equation using Chapman-Enskog expansions 
to Burnett's level, although  restricted to steady states. 
Also the linearized Burnett equations in the SET (in the first 
Enskog approximation) have been used by Alves and Kremer~\cite{alves}
to study light scattering from density fluctuations.
However it would be difficult to compare our approximations with those
from  Chapman-Enskog's method. They are quite different methods indeed.

Furthermore, in order to calculate any physical quantity, we first determine $\chi_{c}$.
The static pressure $p$ is given as $p=\frac{1}{2}{\rm Tr}\mathbf{P}$.

The static pressure $p$ is given as $p=\frac{1}{2}Tr\mathbf{P}$.
This automatically yields that the equation of state is given
by  Eqs.~(\ref{defpressure}) and (\ref{defpcol}).
Then, the relation between $\chi_{c}$ and the static pressure $p$ 
is given by
\begin{equation}\label{dchi}
\chi_{c} = \frac{p/nT - 1}{2\rho_0}
\end{equation}
where the relation: $\delta=\rho_{0}\rm K\!n$ is used.
In general, $\chi_{c}$ is related to the pair
distribution function~\cite{ferziger}.
In the case without external force,
the equation of state at equilibrium is very well approximated
using Henderson's expression~\cite{henderson}
\begin{equation}\label{chi}
\frac{p}{nT}
=\frac{1+\frac{\rho_{0}^{2}}{8}}{(1-\rho_{0})^{2}},
\end{equation}
which is what we use.

In the case with $F\neq 0$, SET does not yield the correct single-particle
equilibrium distribution function, whereas RET do~\cite{hvb}.
 
\section{Heat conduction between parallel plates under 
a steady state condition}\label{sec:heat}

\subsection{Reduced hydrodynamics}

The above hydrodynamics should in principle allow us to study a wide variety
of dynamic problems involving dense gases.  In this section we look at a
quite simple situation in this wide context.  It is the case of a one dimensional
heat conduction case between two parallel plates.

The system consists of  gas between two infinite parallel plates 1 and 2
separated by a distance $L$. The plates have fixed temperatures $T_{1}$ 
and  $T_2$ respectively. A schematic representation in
Fig.~\ref{fig:1} shows the $Y$ axis defined perpendicular to the plates 
while the $X$ axis is placed on plate 1.  
\begin{figure}[htb]
\epsfig{file=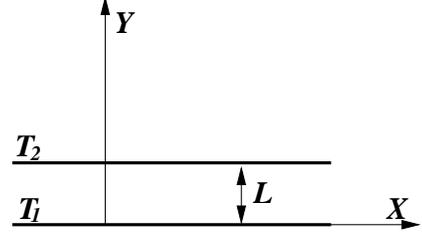,width=5.5cm,angle=0}

\caption{Configuration of the simple heat conduction system 
as described in the text. \label{fig:1}} 
\end{figure} 

In general, under circumstances with a gradient of temperature perpendicular
to the walls, there is a difference between the temperature of the gas by
the plate and the temperature of the plate itself. This is a well-known
effect called thermal slip. For simplicity's sake we neglect such
difference.
 
For this simple system the extended hydrodynamic equations drastically
simplify. The basic concrete equations solved here are the following:
\begin{center}
$P_{yy}(y)\equiv P_{yy}=\hbox{const.}$, $\qquad$ $Q_{y}(y)\equiv Q_{y}=\hbox{const.}$,
\end{center}
\begin{center}
$P_{xy}(y)=p_{xy}(y)=0$,
\end{center}
\begin{widetext}
\begin{equation} \label{EGeq1}
{\it P_{yy}}=-\left[1+\frac{\delta}{\rm K\!n}\chi_{c} n(y) \right] 
                          {\it p_{xx}} \left( y \right) 
+ \left[1+ 2\frac{\delta}{\rm K\!n}\chi_{c} n(y)\right]
           n \left( y \right){\it T} \left( y \right),
\end{equation}
\begin{equation}\label{EGeq2}
Q_{y}= \left[1+\frac{3}{2}\chi_{c} n(y)\frac{\delta}{\rm K\!n}\right]q_{y}^{k}(y)
      -\delta^{2}\frac{2}{\sqrt{\pi}\rm K\!n}\chi_{c} 
                n(y)^{2}\sqrt{T(y)}\frac{dT(y)}{dy},
\end{equation}
\begin{eqnarray}\label{EGeq3}
-\frac{1}{2}\frac{dq_{y}^{k}(y)}{dy}=
&-&\frac{8}{\sqrt{\pi}\rm K\!n}\chi_{c} \sqrt {{\it T} \left( y \right) }
          \left[n \left( y \right) {\it p_{xx}} \left( y \right)
                -\frac {q_{y}^{k}(y)^{2}}{128{\it T}\left( y\right)^{2}}\right]
+\frac{\delta}{4\rm K\!n}\chi_{c} \left[5q_{y}^{k}(y)\frac{dn(y)}{dy}
+3n(y)\frac{dq_{y}^{k}(y)}{dy} \right] 
\nonumber \\
&-& \frac{\delta^{2}}{\sqrt{\pi}\rm K\!n}\chi_{c} n(y)\sqrt{T(y)}
       \left[ 2\frac{dn(y)}{dy}\frac{dT(y)}{dy}
             +\frac{1}{2}\frac{n(y)}{T(y)}\left(\frac{dn(y)}{dy}\right)^{2} 
             +n(y)\frac{d^{2}T(y)}{dy}\right],
\end{eqnarray}
\begin{eqnarray}\label{EGeq4}
&&-T(y)\frac{dp_{xx}(y)}{dy}-2\frac{dp_{xx}(y)}{dy}\frac{dT(y)}{dy}
  +2n(y)T(y)\frac{dp_{xx}(y)}{dY}+\frac{T(y)}{n(y)}p_{xx}(y)\frac{dT(y)}{dy}
  -\frac{p_{xx}(y)}{n(y)}\frac{dp_{xx}(y)}{dy}
\nonumber \\
=&-&\frac{4}{\sqrt{\pi T(y)}\rm K\!n}\left[ n(y)T(y)-p_{xx}(y) \right]
q_{y}^{k}(y)  
\nonumber \\
&+& \frac{\delta}{\rm K\!n}\chi_{c} 
\left[  \frac{7}{2}n(y){\it T} \left( y \right)\frac{dp_{xx}(y)}{dy} 
       +\frac{7}{2}T \left( Y \right)p_{xx}(y)\frac{dn(y)}{dY}
       +2{\it p_{xx}} \left( Y \right)\frac{dT(y)}{dy}
       -8n(y)T(y)^{2}\frac{dn(y)}{dy}
       -7n \left( Y \right)^{2}T(y)\frac{dT(y)}{dy}\right].
\end{eqnarray}
\end{widetext}

\subsection{Functional expression of the pressure}\label{sec:results}

All the results we describe in what follows were obtained using  perturbation
methods choosing $T_2>T_1$ and using $\epsilon= (T_{2}-T_{1})/T_{1}$ as
perturbation parameter. We solve the system of equations up to $\epsilon^{6}$
and try to express the nonequilibrium pressure $P_{yy}$ in terms of the density, 
temperature and the heat flux $Q_{y}$. Note that here $P_{yy}$ is constant.
We  manage to do this up to third order in  $\epsilon$, which corresponds to  
situations not very far from equilibrium. The functional expression for the 
nonequilibrium pressure $P_{yy}$ turns out to be
\begin{widetext}
\begin{equation}\label{functional}
P_{yy}[n(y),T(y),Q_{y}]=n(y)T(y)\left[1+2\chi_{c} \frac{\delta}{\rm K\!n}n(y)\right]
          \left[1+\lambda_{p}\frac{Q_{y}^{2}}{n(y)^2T(y)^3}\Lambda[n(y),T(y)]\right]
\end{equation}
where $\lambda_{p}$ is the following constant
\begin{eqnarray}\label{lambdap}
\lambda_{p}=&-&\frac{1}{128}+\frac{7}{16}\chi_{c} \frac{\delta}{\rm K\!n}
                             -\frac{1}{128}\left[\frac{9}{\pi}+2253\right]
                                      \chi_{c}^{2}\left[\frac{\delta}{\rm K\!n}\right]^{2}
\nonumber \\
            &+&\left[ \frac{7}{32}\chi_{c} \frac{\delta}{\rm K\!n}
                     -\frac{1}{128}\left[\frac{9}{\pi}+2309\right]
                             \chi_{c}^{2}\left[\frac{\delta}{\rm K\!n}^{2}\right]\right]\epsilon
\nonumber \\
            &+&\left[ \left[-\frac{7}{384}+\frac{7\pi \rm K\!n^{2}}{16384\chi_{c}^{2}}\right]
                                                                 \chi_{c} \frac{\delta}{\rm K\!n}
                     -\frac{1}{128}\left[ \frac{2356}{6}+\frac{3}{2\pi}
                                         -\frac{3+751\pi \rm K\!n^{2}}{256\chi_{c}^{2}}\right]
                             \chi_{c}^{2}\left[\frac{\delta}{\rm K\!n}^{2}\right]\right]\epsilon^{2}
\nonumber \\
            &+&\left[ \left[\frac{7}{768}+\frac{7\pi \rm K\!n^{2}}{16384\chi_{c}^{2}}\right]
                                                                 \chi_{c} \frac{\delta}{\rm K\!n}
                     -\frac{1}{128}\left[\frac{3+359\pi}{256}\right]\delta^{2}\right]\epsilon^{3}
\end{eqnarray}
and where
\begin{equation}\label{Lambda}
\Lambda[n(y),T(y)]=1+n(y)(1+56T(y))\chi_{c} \frac{\delta}{\rm K\!n} 
                    +\left[-\frac{32+305\pi}{4\pi}+784T(y)
                                                  +\frac{3(257\pi-3)}{\pi}T(y)^{2}\right].
\end{equation}
\end{widetext}
Substituting $\delta=0$ in 
Eq.~(\ref{EGeq1}) yields the functional expression 
for the pressure $P_{yy}$, 
\begin{equation}\label{fnlbg}
P_{yy}(n,T,Q_{y})=n(y)T(y)\left[1
 -\frac{1}{128}\frac{\mathbf{Q_{y}^2}}{n(y)^{2}T(y)^{3}}\right].
\end{equation}
Therefore Eq.~(\ref{functional}) is a natural extension of Eq.~(\ref{fnlbg}).
For hard spheres the expression which correspond to Eq. (\ref{fnlbg}) was 
already given in~\cite{grad}. 
For hard disks Eq. (\ref{fnlbg}) is given for the first time 
but  it can be obtained from the appendix  in~\cite{pc}, where a gas of 
{\em inelastic} disks is studied.

\subsection{Comment}\label{sec:comment}

The steady state thermodynamics (SST) proposed by Sasa and
Tasaki~\cite{sasa} assumes that all nonequilibrium variables and
thermodynamic functions can be expressed in terms of the
density, temperature and the steady state fluxes (heat flux in the case
of our example).  We found that using Grad's moment method~\cite{grad,
grad58}, this condition is satisfied in the simple heat conductive
case for the dilute and dense gases which obey Boltzmann's or Enskog's
kinetic equations in the range of the approximation introduced
in Sec.~\ref{sec:simplify} if the system is not too far from equilibrium.
Unfortunately, because kinetic theory defines only the internal energy
density, $u(y) \propto T(y)$, and this quantity is interpreted
as the temperature, it is impossible, within the framework of kinetic
theory alone, to construct nonequilibrium thermodynamic functions like the
Helmholtz energy. Using Boltzmann's entropy: $s=-f \ln f$ ($s$ is the 
entropy density), one can define the entropy and a 
temperature through $(\partial s/\partial u)=1/\theta$ where $\theta$ is the
nonequilibrium temperature~\cite{jou} in what is called {\em extended
irreversible thermodynamics} (EIT)~\cite{jou}. Furthermore one could define
the nonequilibrium Helmholtz free energy and chemical potential, 
although we have not seen anyone developing such a formalism. 
However within kinetic theory it is not clear
which is the difference between the temperature and the internal energy
density.  Even if the entropy of the system were given it is not possible to
determine Helmholtz free energy and therefore it is impossible to determine
the chemical potential. Using an assumption of SST~\cite{sasa} that the
Maxwell relation is satisfied by the nonequilibrium thermodynamic variables,
then from Eq.~(\ref{functional}) a chemical potential follows without having
to determine Helmholtz's free energy, although this chemical potential still
has an undetermined part which depends on the heat flux and temperature. 
In a following  article~\cite{cpotential} we propose a method to get the
chemical potential partially using the framework of SST~\cite{sasa}. 
The undetermined part is measured experimentally. 
 By this method, one can determine the value of 
the chemical potential at the interface between equilibrium 
and nonequilibrium sides.
In other words, we will propose an experimental method for determining
the values of the nonequilibrium variables at the equilibrium-nonequilibrium interface 
following the Sasa and Tasaki scheme.
Once a chemical potential is given it is possible to completely express  
the condition that the perfect $\mu$ wall must have~\cite{sasa} 
and to discuss about the prediction of ~\cite{sasa} beyond any doubt.

\section{Summary and final remarks}\label{sec:summ} 

In this article we have given in detail the hydrodynamic
equations for a bi-dimensional dense gas of hard disks that
follow from Enskog's equation using Grad's moment expansion
method.  Bearing in mind that without a strategy to
simplify the calculations no manageable hydrodynamics can
be obtained we have introduced in Sec.~\ref{sec:simplify} a
simplifying scheme that leads to a hydrodynamics which is
more complete than the one that would follow from a linear
approximation scheme.  As far as we know this is the first
time that Grad's method used to obtain extended
hydrodynmaic equations from Enskog's equation has been
published beyond the a linear approximation in two or three
dimensions.

We have applied this hydrodynamic equations to discuss a simple but subtle
1D case of heat conduction. The example has the virtue of giving the
opportunity to discuss the adequacy of the starting point of
SST~\cite{sasa}.  We have shown that this simple nonquilibrium steady state
system, not too far from equilibrium, behaves as if nonequilibrium
thermodynamic variables could be expressed in a functional form.  Hence, it
may be possible to construct a nonequilibrium thermodynamics without any
information of the microscopic dynamics of the system. First, we have shown
this for the dense hard disk system. It is well known that for the
hydrodynamics obtained from Boltzmann's equation using Grad's method this is
correct~\cite{grad, grad58}.

\acknowledgments
{ The authors  express their gratitude to  D. Risso and R. Soto for
many helpful conversations.  One of us (H.U.) thanks the scholarship from
{\em Mecesup UCh 0008 project} while the other (P.C.) thanks his partial
finance from {\em Fondecyt} research grant 1030993 and {\em Fondap}
grant 11980002.
}

\end{document}